# Alternating Current Electrothermal Flow for Energy Efficient Thermal Management of Microprocessor Hot Spots

Golak Kunti, Jayabrata Dhar, Anandaroop Bhattacharya, Suman Chakraborty[a]

Department of Mechanical Engineering, Indian Institute of Technology Kharagpur, Kharagpur, West Bengal - 721302, India

[a]*E-mail address of corresponding author:* suman@mech.iitkgp.ernet.in

**ABSTRACT**

The paper presents numerical results on an innovative concept of an efficient, site-specific cooling technology for microprocessor hot spots using Alternating Current Electrothermal Flow (ACET). ACET deploys an electrokinetic transport mechanism without requiring any external prime mover and can be shown to be highly effective in reducing hot spot temperatures below their allowable limits. The proposed technique leverages the heat source(s) itself to drive the fluid in the ACET cooling mechanism, thereby making this a highly energy efficient active technique. Our parametric analyses present the optimal range of fluid properties, viz. electrical conductivity where the cooling efficiency is maximum. Further, the impact of geometrical parameters as well as input voltages has been characterized. Our results show a reduction of hot spot temperature by more than 30% for flow through a channel of 600 microns under appropriate conditions at an applied AC voltage of 10-12 V. Comparison with other micro-cooler technologies show the proposed technique to be more energy efficient in its range of operation. These results establish the potential of the ACET cooling and can open up a new avenue to be explored for thermal management of miniaturized devices and systems.

# 1. Introduction

Thermal design of microelectronic equipment plays a crucial role in the electronic industry [1–3]. In today's world, microprocessors have emerged as a backbone of the electronics industry owing to reduction in cost of electronic packages through technological advances. After Moore's prediction of exponential growth in production of integrated circuits, a self-fulfilling prophecy was developed by Dennard scaling rules [4]. Microprocessors are widely used in data centers for development of Information Technology (IT). As a consequence, energy consumption rate increases rapidly. The annual electricity consumption of data centers was 238 billion kWh in 2010 [5]. The statistics of the International Technology Roadmap for Semiconductor (ITRS) predicts generated power density of an electronic device can reach up to 100 W/cm$^2$ by 2020 [6]. The required energy for running the data centers disturbs today's ecological balance. An efficient thermal management system is required to reduce the bills of electricity of these data centers [7].

For the safety and proper functionality of microelectronic components, it is important to remove the heat generated within the electronic component so that the component temperatures are kept within the safe operating limits. Due to the recent technological advancements, microelectronic devices have to efficiently control numerous functions simultaneously, resulting in increased heat dissipation levels within a reduced confinement [8]. Depending on the design of the microprocessor, the heat flux is mostly non-uniform with concentrated regions of high values, which results in the formation of localized hot spots [9]. Thus, effectiveness and reliability in heat removal process from the processor/device in general and hot spots in particular, are the pivotal issues of a compact thermal management system [10,11].

Most of the data centers uses air-cooling system for proper of the server. Air has very poor capacity of transporting heat and large power is required to transport air [12]. However, to overcome the drawbacks of air-cooling technology, the present thermal management systems employ other techniques such as liquid cooling [13], phase change material (PCM) [14,15], two-phase cooling [16,12] and use of nanofluids [17]. PCM relies on absorbing latent heat and releasing thermal energy during off-peak and peak load period, respectively. Two-phase cooling process employs micro-evaporator cold plates which are attached with CPUs and are much efficient than conventional air-cooling.

Tuckerman first introduced microchannel heat sink to improve the heat transfer efficiency [18]. On the other hand, Marschewski et al. reported efficient cooling of 3D electronic chip by triggering helicoidally fluid motion using Herringbone microstructures [19]. To circumvent the problem of decreased heat transfer rate along straight channels, secondary flow to enhance fluid mixing was explored and reported [20]. Over the past decade, further enhancement in thermal performance through introduction of nanoparticles in the fluid (nanofluids) for flow through microchannels have been reported by several research groups [17] Furthermore, several optimization methods, namely Taguchi method, non-sorted genetic algorithm (NSGA), multi-objective optimization were developed to improve heat transfer performance and to reduce pressure drop [21–23]. In spite of some inherent benefits of the above methodologies of thermal management, these are not suitable at reduced volume and smaller dimension.

An important issue of microelectronic devices is concentrated heat generation (power dissipation) at specific locations. High power dissipation in the core regions generates micro-hotspots locally. This calls for a shift from traditional microchannel designs with uniform flow to some novel ideas that enhance the heat transfer rates locally at the hot spots [24]. While passive cooling approaches including hotspot-targeted embedded liquid cooling or microgaps with variable pin fin clustering can help at localized hotspot mitigation [1,24], their effectiveness is limited due to some inherent disadvantages.

On the other hand, thermoelectric cooling (TEC) [25–27] and electrowetting [28,29] offer active methods of hot spot thermal management that require an external power source for operation. TEC suffer from the drawbacks of low energy conversion efficiency, high cost and introduction of additional thermal resistances in the heat flow path [30,31]. Electrowetting on Dielectric (EWOD), on the other hand, suffer from high voltage requirements for required pumping power, thereby limiting its applicability[32,33].

To overcome the limitations stated above, we propose an energy efficient hotspot targeted microscale cooling actuated by alternating current electrothermal forces. To the best of our knowledge, electrothermally based microcooling of hot spots has not been studied so far. Electrothermally driven fluid flow arises in presence of local variation of electrical conductivity and permittivity due to presence of temperature gradients in the conducting fluid [34–36]. The source of the thermal energy may be external, such as strong illumination [37] or maybe the Joule heating generated by application of electric field [38,39]. In our application, pertaining to thermal management of localized hot spots, we propose to use the hot spots themselves as the source of thermal energy. ACET mechanisms, employed effectively for fluid pumping [40–44] and mixing [45–48], can thus be effectively utilized for hot spot cooling. This concept is highly energy saving and presents a unique scenario where the source of the problem (hot spot) is used as the solution (driver for flow).

**2. Numerical methods**

Electrothermal forces originate from the movement of the induced charges generated by local gradient of the electrical conductivity and permittivity which in turn depends on established temperature gradients in the bulk fluid. Fig. 1(a) shows the electrothermal mechanism schematically, whereas Fig. 1(b) shows the physical system which consists of three hot spots located at the middle of the micro-cooler. Six pairs of electrodes are oriented on the floor (three to the left of the hot spot and other three to its right) to drive the coolant. The height and length of the channel are $H$ and $L$, respectively where $L$ is held constant at 500 μ-m while $H$ is be varied to determine its optimum value for maximum heat transfer rate. The dimensions of the various length for the electrode pairs are as follows: $d_1 = g = 6\,\mu m, d_2 = 36\,\mu m$ and $e = 22\,\mu m$ (see Fig 1(b)). The length of the hot spot is $h = 20\,\mu m$ and gap between hot spots is $s = 10\,\mu m$. An external AC electric field is applied across the electrodes pairs located on the channel floor. An energy conversion cascade diagram is shown in Fig. 1(c) depicting the conversion of thermal energy from the hot spots and electrical energy from the AC signal to kinetic energy that eventually leads to convective heat dissipation.

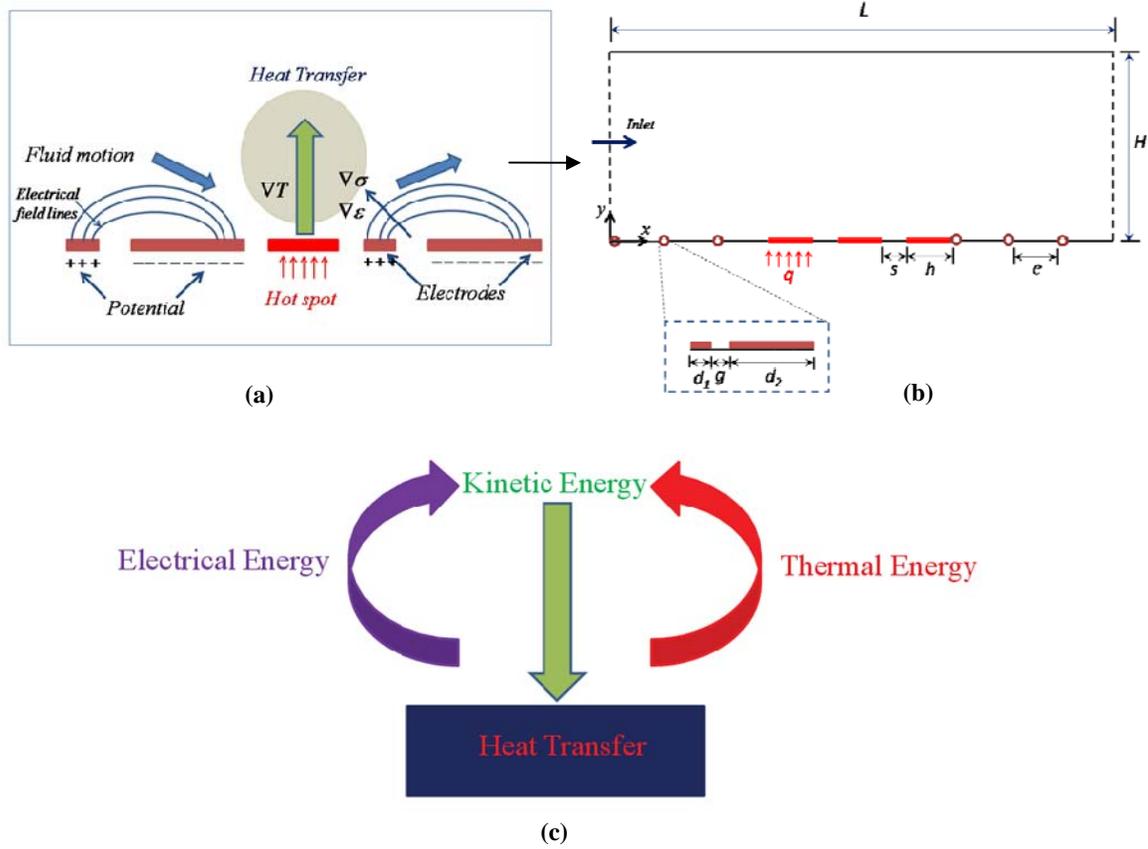

**Fig. 1**. (a) Electrothermal mechanism in presence of hot spots. (b) Schematic representation of the physical system. Electrodes are marked (shown in the dotted box) to show their position relative to the wall. Three hot spots are placed at the middle of the microchannel floor (shown by red color). (c) cascade diagram of energy conversion and heat transfer process through electrothermal mechanism. The thermal energy of the hot spots and externally applied electrical energy combinedly generates kinetic energy to reduce the temperature of the hot spots.

In this subsection, we describe the governing equations involved with the ACET fluid flow and heat transfer characteristics pertaining to our proposed hot spot cooling mechanism. Central to this theory lies the simultaneous coupling of electric field, temperature field and flow field. Under the condition of quasi-electrostatic field (i.e., negligible magnetic field effect [49,50]), the electrostatic field **E** = -$\nabla \varphi$ is governed by the equation [51,52]:

$$\nabla \cdot (\nabla \varphi) = 0, \tag{1}$$

where $\varphi$ is the electrostatic potential and $\varepsilon$ is the fluid permittivity. As mentioned before, ACET forces are generated from the variation of permittivity and electrical conductivity due to a non-uniform thermal field of the coolant. The expression for the dependence of permittivity and electrical conductivity ($\sigma$) with temperature can be written as

$$\varepsilon(T) = \varepsilon_0(T_0)(1+\alpha T),$$
$$\sigma(T) = \sigma_0(T_0)(1+\beta T), \qquad (2)$$

where $T_0$ is the reference temperature. For aqueous solution, $\alpha = -0.004\,\mathrm{K}^{-1}$ and $\beta = 0.02\,\mathrm{K}^{-1}$ [34].

While a small amount of Joule heat arises from the AC electric field, the primary sources of heat are the hot spots that result in the temperature gradients across the microchannel. The temperature distribution can be captured by the following energy equation:

$$\rho C_p \frac{DT}{Dt} = \nabla \cdot (k\nabla T) + \sigma|\mathbf{E}|^2, \qquad (3)$$

where $T, \rho, C_p, \mathbf{V}$ and $k$ are the temperature, density, specific heat, flow velocity vector and thermal conductivity of the fluid, respectively. The term $\sigma|\mathbf{E}|^2$ is the Joule heat generated from the electrical field within the fluid domain while $D/Dt$ represents the material derivative.

Assuming laminar, incompressible flow, the Navier-Stokes equation, which governs ACET fluid flow, is expressed by

$$\nabla \cdot \mathbf{V} = 0, \qquad (4)$$

$$\rho \frac{D\mathbf{V}}{Dt} = -\nabla p + \nabla \cdot \left[\mu(\nabla \mathbf{V} + \nabla \mathbf{V}^T)\right] + \mathbf{F}_\mathbf{E}, \qquad (5)$$

where $\mu$ is the fluid viscosity and $p$ is the pressure. Here the fluid motion is developed by the body forces ($\mathbf{F}_\mathbf{E}$) originating from electrothermal effects. The time averaged electrothermal force $\mathbf{F}_\mathbf{E}$ has the form[34,49]:

$$\mathbf{F}_\mathbf{E} = -\frac{1}{2}\left[\left(\frac{\nabla\sigma}{\sigma} - \frac{\nabla\varepsilon}{\varepsilon}\right)\cdot \mathbf{E}\frac{\varepsilon\mathbf{E}}{1+(\omega\tau)^2} + \frac{1}{2}|\mathbf{E}|^2 \nabla\varepsilon\right], \qquad (6)$$

where $\tau = \varepsilon/\sigma$ denotes the charge relaxation time of the electrolyte and $\omega = 2\pi f$ is the angular frequency of the AC potential. Since, $\varepsilon$ and $\sigma$ are functions of temperature only we can write $\nabla\varepsilon = (\partial\varepsilon/\partial T)\nabla T$ and $\nabla\sigma = (\partial\sigma/\partial T)\nabla T$. We have neglected the electrostriction force, $\nabla \rho (\partial \varepsilon / \partial \rho) \mathbf{E} \cdot \mathbf{E}$ in the body force terms under the consideration of negligible permittivity variations with fluid density. An AC potential ($\pm V_{rms}$) at a frequency of 100 kHz, with 180° phase difference, is imposed on the electrode pairs located on the channel floor. Other boundaries are electrically insulated. For temperature distribution, inflow boundary condition is used as $-\mathbf{n}.(-k\nabla T) = \rho(\Delta h_{in} - \Delta h_{ext})\mathbf{V}.\mathbf{n}$, $\Delta h_{in} - \Delta h_{ext} = \int_{T_{ext}}^{T_i} C_p dT$, where $(\Delta h_{in} - \Delta h_{ext})$ is the enthalpy difference of the fluid between the inlet and external ambient temperatures, $T_i$ is the inlet temperature of the fluid, $\mathbf{n}$ is the unit normal vector (directed along x direction). Since the hot spots and AC field result in temperature fields throughout the channel, it is expected that the inlet fluid temperature will be higher that the external ambient. The external temperature in our model is set to $T_{ext} = 298K$ while the heat flux at the hot spot is specified as $10^6$ W/m² or 100 W/cm². At the top wall, an effective heat transfer coefficient of $h_c$ (ambient temperature is 298K) is imposed. At the outlet of the channel, $\partial T / \partial x = 0$, while the other boundaries are thermally insulated. For velocity field, no slip is set on the solid boundaries and zero pressure is imposed on the inlet and outlet, since no other external forces are present.

To quantify the performance of the ACET micro-cooler, we define an index $\varepsilon_{ACET}$ to denote the micro-cooler effectiveness:

$$\varepsilon_{ACET} = \frac{T_{max,no\ cooler} - T_{max}}{T_{max,no\ cooler} - T_{ext}}. \tag{7}$$

where $T_{max,no\ cooler}$ and $T_{max}$ are the maximum temperature in the domain when the ACET cooler is deactivated and activated, respectively. $T_{ext}$ is the external temperature of the fluid before entering the channel. As evident from the expression, $\varepsilon_{ACET} = 1$ denotes the theoretical maximum where the hot spot is cooled down to the external ambient temperature. On the other hand $\varepsilon_{ACET} = 0$ (i.e., $T_{max} = T_{max,no\ cooler}$) implies that the micro-cooler is completely ineffective in reducing the hot spot temperature.

In our analysis, the working fluid was chosen as KCl electrolytic solution having the following properties:

$\rho = 1000 \, \text{kg/m}^3$, $C_p = 4{,}184 \, \text{J/kgK}$, $\mu = 0.00108 \, \text{Pa s}$, $\varepsilon = \varepsilon_0 \varepsilon_r = 7.08 \times 10^{-10} \, \text{C/Vm}$. Here $\varepsilon_0$ is permittivity of vacuum and $\varepsilon_r$ is relative permittivity of the liquid (i.e., $\varepsilon_0 = 8.85 \times 10^{-12} \, \text{C/Vm}, \varepsilon_r = 80$), respectively. The geometrical and physical parameters stated above are constant for all the results and the values of some other parameters are parametrically varied in the subsequent studies.

The numerical investigations were studied using COMSOL, a Finite Element Method based commercial Multi-physics software package. In our solution scheme, Eq. (1) is solved first to distribute the electric field over the domain. Using this potential distribution the temperature and velocity fields are next solved simultaneously (Eq. (3)-(5)). The solutions from the above coupled fields are next used to obtain the ACET body forces (Eq.(6)). Finally, the resultant ACET forces are employed to drive the fluid over the hot spots for cooling. In the numerical model, a finer mesh is used near the electrodes and hot spots to capture the large gradients in electric, thermal and velocity fields. It is important to note that in order to satisfy the numerical convergence criteria a very low value of velocity (order of $10^{-14}$) is set at the inlet of the channel to obtain the maximum temperature when the cooler is deactivated.

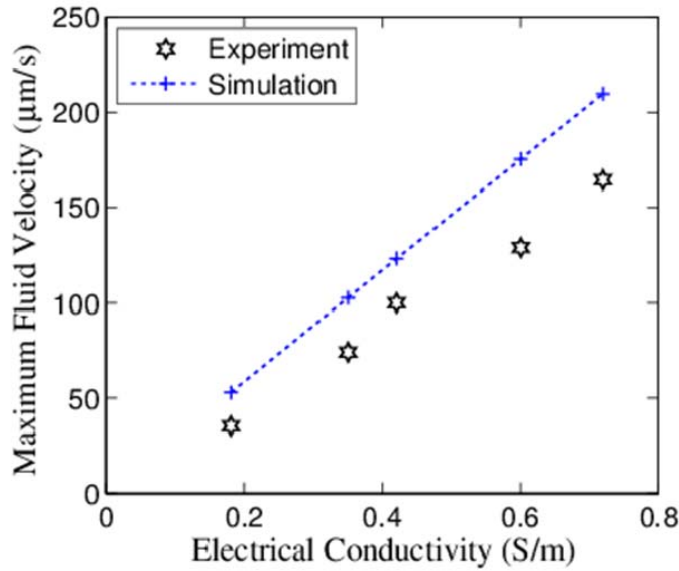

**Fig. 2**. Model benchmarking: Variation in maximum fluid velocity with solution concentration. Dotted line shows the numerically simulated results whereas star symbols depict experimental data. Numerically obtained trend of variation of velocity matches well with experimental data reported in Vafaie et al.[53].

After systematic grid independence test the numerical modeling was extensively benchmarked with the experimental results of Vafaie et al.[53], whose study was based on investigations of bi-directional fluid flow using electrothermal kinetics. The comparison is shown in Fig. 2 in terms variation of maximum fluid velocity as a function of solution concentration. It is evident that with increasing concentration Joule heating increases, thus leading to larger temperature gradient. This is turn results in stronger electrothermal forces and consequently higher fluid velocities. It is seen that this trend as well as the slope matches well between our model predictions and the experimental results of Vafaie et al.[53]. The deviation in values can be attributed to the ideal considerations adopted in the numerical modeling where the effects of electrothermal reaction, buoyancy forces, etc are not considered, resulting in over prediction of maximum flow velocities.

## 3. Results and discussions

From the general concept of electrothermal effect, it is expected that increasing voltage on the electrodes will increase pumping velocity, and hence, reduce the hot spot temperature. However, within the micro-cooler, electric field in conjunction with high electrical conductivity generates significant Joule heat. Given a specific electrical conductivity, Joule heat is proportional to square of the electric field strength (Joule heat $\sim E^2$). Therefore, the volumetric heat generation increases with square of the applied potential. Thus, on one hand, increasing voltage increases the flow velocity and flow vortices leading to enhanced heat transfer, while, on the other hand, the induced Joule heat increases the overall temperature of the system. The net effect of these two opposing trends on cooling effect of hot spots is depicted in Fig. 3 for two realistic values of the electrical conductivity (Fig. 3(a) $\sigma = 2, 10 \, \text{ms/m}$) and height (Fig. 3(b) $H = 350, 600 \, \mu\text{m}$). The other relevant parameters are shown in the caption. Further, from Fig. 3(a) it is clear that the maximum temperature of the hot spots decreases sharply at lower range of potential while the drop is more gradual at higher values.. The cause, as described above, can be attributed to higher Joule heating at increased voltages resulting in heating up of the bulk fluid (coolant) .The variation of the cooler effectiveness, $\varepsilon_{\text{ACET}}$, is consistent with this observation (see the inset figure of Fig. 3(a)). For the range of applied potential $\varphi = 10 - 12 \, \text{V}$, the cooling effectiveness asymptotically reaches a maximum value above which any further increase in potential does not have significant effect in enhancement of convective heat transfer.

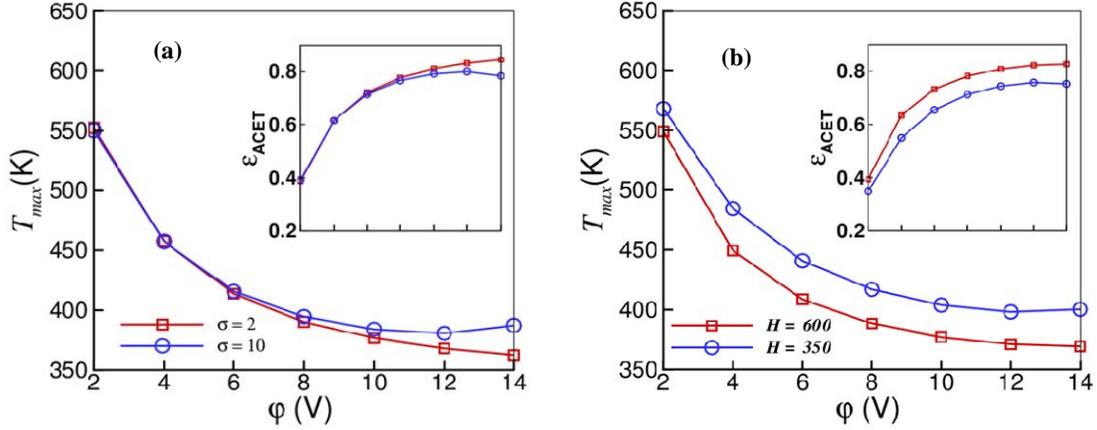

**Fig. 3.** Effect of applied potential ($\varphi$) on the variation of maximum temperature of hot spots and ACET effectiveness ($\varepsilon_{ACET}$) (inset) as a function of $\varphi$. we consider two different cases: (a) two different values of electrical conductivity $\sigma = 2$ and $10\,\text{mS/m}$, other parameters used in the results are thermal conductivity $k = 0.6\,\text{W/mK}$, heat transfer coefficient $h_c = 500\,\text{W/m}^2\text{K}$, channel height $H = 600\,\mu\text{m}$ and (b) two different values of channel height $H = 350$ and $600\,\mu\text{m}$, other parameters used in the results are electrical conductivity $\sigma = 5\,\text{mS/m}$, thermal conductivity $k = 0.7\,\text{W/mK}$, heat transfer coefficient $h_c = 500\,\text{W/m}^2\text{K}$.

Fig. 3(b) shows the variation of maximum temperature of the hot spot with applied potential for heights $H = 350$ and $600\,\mu\text{m}$. The secondary flows or vortices enhance convective heat transfer due to mixing of fluid over the hot spots. However, too many vortices result in low flow rate, thereby, adversely affecting the heat transfer rate. At higher channel heights, the overall flow resistance is low and the relative contribution of the vortices is thus lower. From the figure, it is evident that the peak temperature of the hot spots is strongly affected by height of the microchannel. For $H = 350\,\mu\text{m}$, the peak temperature hardly goes below 400K though the applied potential is in the range 10-12V. On the other hand, for $H = 600\,\mu\text{m}$, the peak temperature of the hot spot falls below 370K with application of potential 10-12V. Corresponding plot of effectiveness of heat transfer with potential (inset Fig. 3(b) also depicts the same feature. Across all the data for applied potential, there is a large change in effectiveness with variations in channel height. Therefore, keeping other parameters constant, the channel height is likely to have significant effect on electrothermal effectiveness. From the Fig 3(a) and 3(b), it is clear that our designed micro-cooler is best suited in the potential range of 10-12V.

Apart from applied potential for reliable and safe performance of the micro-cooler, electrical properties of the fluid and geometrical properties of the system that affect the flow behavior and heat transfer characteristics also demand a closer look. To this end, we investigates the effect of electrical conductivity for different channel height and effect of channel height for different electrical conductivities.

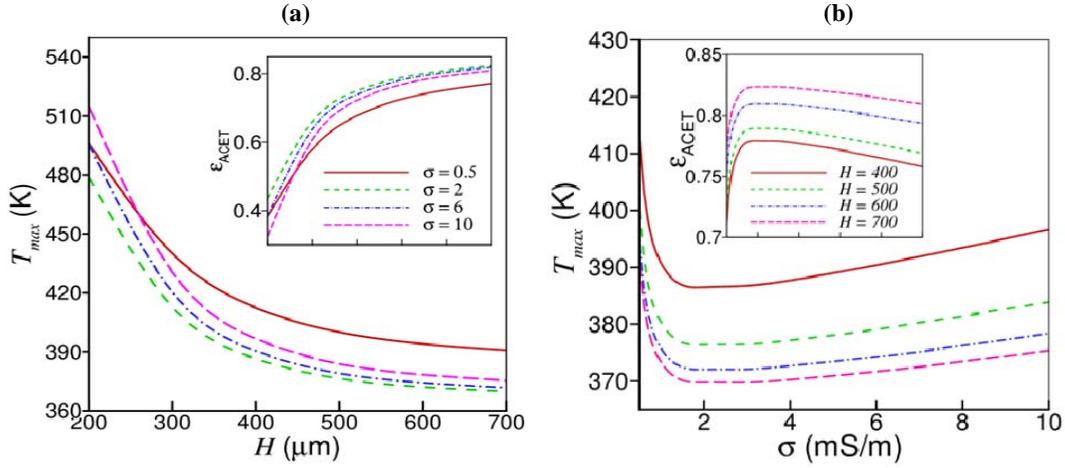

**Fig. 4.** Effect of electrical conductivity ($\sigma$) and channel height ($H$) on the variation of maximum temperature of hot spots and ACET effectiveness ($\varepsilon_{ACET}$) (inset). (a) four different electrical conductivity $\sigma = 0.5, 2, 6 \text{ and } 10 \,\mu\text{m}$, are chosen to analysis the influence of channel height. The other parameters used in the analysis are AC voltage $V_{rms} = 10 \,\text{Volt}$, thermal conductivity. $k = 0.7 \,\text{W/mK}$, heat transfer coefficient $h_c = 500 \,\text{W/m}^2\text{K}$. (b) four different channel height $H = 400, 500, 600 \text{ and } 700 \,\mu\text{m}$, are taken to study the effect of electrical conductivity. The other relevant parameters are AC voltage $V_{rms} = 10 \,\text{Volt}$, thermal conductivity $k = 0.7 \,\text{W/mK}$ heat transfer coefficient $h_c = 500 \,\text{W/m}^2\text{K}$.

From Fig. 4(a), it is seen that at lower height (typically H=200 $\mu\text{m}$), the maximum temperature is high. However, the peak temperature sharply decreases up to H=400 $\mu\text{m}$ and finally goes below 378K. Beyond a threshold height (H=600 $\mu\text{m}$), any further increase in height does not affect the maximum temperature significantly. This is further corroborated by the inset of the Fig. 4(a), where the electrothermal effectiveness saturates beyond H=600 $\mu\text{m}$. The above trends can be attributed to the fact that at lower channel height, formation of recirculating vortices blocks the flow passage and retards longitudinal flow. Further, a proper temperature gradient does not get established for channels with lower heights, which is essential for higher advective flow rates. Although flow vortices enhance heat transfer, less advection due to large vortices in flow field results in reduced convective heat transfer rate. Furthermore, as electrical conductivity is varied from $\sigma = 0.5 \,\text{mS/m}$ to 10 mS/m, the peak temperature of the hot spots drop to below 380K beyond 2mS/m.. A closer look at the plots reveal that the minimum peak temperature as well as the highest effectiveness occurs at $\sigma = 2 \,\text{mS/m}$, beyond which further increase in electrical conductivity is actually detrimental to the cooling effectiveness. This is consistent with the observation discussed previously (in Fig. 3(a)). Fig. 4(b) demonstrates that for all values of channel height, the maximum temperature of the hot spots first decreases sharply, experiencing a minimum near $\sigma = 2$ mS/m, and then gently increases. The important insight from the plot is that at very low conductivity, the electrical field is insufficient to pump the fluid. On the other hand, at high conductivity, significant Joule heat induces, which enhances the fluid temperature. This is consistent with the prior observations and explanations. From the electrothermal efficiency plot (inset of Fig. 4(b)), it is also clear that the effectiveness experiences a maximum at a particular optimal value of the electrical conductivity.

The results from the above section can thus be summarized as: (a) for a given potential, there exists a solution electrical conductivity irrespective of channel height where we observe a minima in the peak temperature. (b) beyond a critical channel height, electrothermal effectiveness does not increase significantly, which for configuration, corresponded to $H=600\,\mu$m. The details of thermal and flow field for optimized operating conditions ($H = 600\,\mu$m, $\sigma = 2$mS/m) are investigated in the supplementary material (Fig. S1).

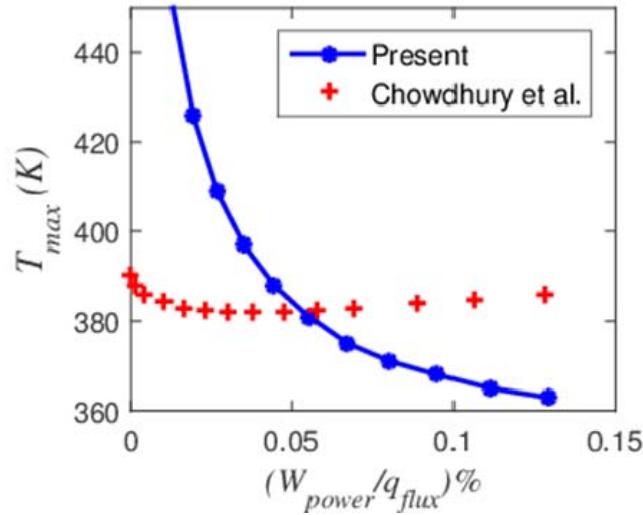

**Fig. 5.** Comparison of cooling performance of present study against the data of Chowdhury et al.[30]. The parameters used in our study are: height $H = 600\,\mu$m, electrical conductivity $\sigma = 2$mS/m, thermal conductivity $k = 0.6$ W/mK and heat transfer coefficient $h_c = 500$ W/m$^2$K. The ACET micro-cooler is much efficient than thermoelectric cooling above a threshold operating power input.

Finally, we compare the cooling performance of the proposed arrangement with the study reported from Intel on Bismuth Telluride based thin film thermoelectric coolers that was demonstrated for on-demand cooling of microprocessor hot spots through integration with the integrated heat spreader [33]. Fig. 5 compares the reported values of maximum hot spot/chip temperature with our proposed technology for various operating power budgets. The operating power is normalized with respect to heat flux applied at the hot spots. For our ACET micro-cooler, the optimized values of $H = 600 \mu m$ and $\sigma = 2 mS/m$ were chosen. The other relevant parameters are provided in the caption. It is observed that below a critical operating power, the supper-lattice based thin film TEC is more efficient compared to our ACET micro-cooler while the trend reverses at higher values. This is due to the higher thermal conductivity of the TEC that replaces part of the thermal interface material in the heat flow path resulting in lower hot spot temperatures. On the other hand, the electrothermal forces at lower applied voltages are not strong enough to get sufficient fluid flow and vortices. However, beyond a threshold limit, stronger ACET forces kick in resulting in enhanced cooling.

It may be noted that similar to TEC, the ACET micro-cooler also holds the advantage of on-demand actuation. This implies that the micro-cooler can be actuated only beyond a threshold temperature of the hot spots, which in turn, depend(s) on the work load on the microprocessor. This feature obviates the need for continuous operation and makes it very attractive from the energy consumption point of view.

The optimized operating conditions of ACET cooling result in greater advantages over electrowetting-on-dielectric (EWOD) as well due to much lower voltage requirements [29]. We also compared the heat transfer characteristics of two different actuation mechanisms: pressure driven flow and ACET driven flow. Results reveal that ACET driven cooling is more efficient than pressure driven actuation when compared with respect to pumping power (see supplementary material (Fig. S2)). Finally, we examined the variation of the peak temperature with heat transfer coefficient experienced at the top wall instead of insulated boundary conditions as well as sensitivity to thermal conductivity (see supplementary material (Fig. S3)). Results reveal that the peak temperatures are largely insensitive to these parameters in our range of operation.

**4. Conclusions**

In summary, numerical results are presented for targeted hot spot cooling of microelectronic processors, by employing a novel strategy based on alternating current electrothermal flow. The uniqueness of our strategy is based on the fact that accumulated heat of hot spots itself triggers the fluid flow which, in turn, is conducive for localized cooling. In other words, the source of the problem is leveraged to serve as the solution. Results show the technique to be energy efficient, amenable to on-demand operation, and leading to better cooling performance as compared to other reported strategies, over appropriate operating regimes. We believe that the results of our study and inferences drawn thereof are compelling enough to warrant further investigation of this new technology for localized hot spot cooling of microprocessors in electronic devices and systems.